\documentclass[preprint2]{aastex}

\usepackage{amsmath,amssymb,graphicx,natbib}

\shorttitle{Multiple Transiting Planet Candidates}
\shortauthors{Steffen \textit{et al.}}

\slugcomment{FERMILAB-PUB-10-198-A}

\begin{document}

\title{Five \textit{Kepler} target stars that show multiple transiting exoplanet candidates}
\author{
Jason H. Steffen$^1$, 
Natalie M. Batalha$^2$, 
William J. Borucki$^3$,
Lars A. Buchhave$^{4,5}$, 
Douglas A. Caldwell$^{3,11}$,
William D. Cochran$^6$, 
Michael Endl$^6$, 
Daniel C. Fabrycky$^4$, 
Fran\c cois Fressin$^4$, 
Eric B. Ford$^7$, 
Jonathan J. Fortney$^8$, 
Michael J. Haas$^3$, 
Matthew J. Holman$^4$, 
Steve B. Howell$^9$, 
Howard Isaacson$^{10}$, 
Jon M. Jenkins$^{3,11}$,
David Koch$^3$, 
David W. Latham$^4$, 
Jack J. Lissauer$^3$, 
Althea V. Moorhead$^7$, 
Robert C. Morehead$^7$, 
Geoffrey Marcy$^{10}$, 
Phillip J. MacQueen$^6$, 
Samuel N. Quinn$^4$, 
Darin Ragozzine$^4$, 
Jason F. Rowe$^3$, 
Dimitar D. Sasselov$^4$
Sara Seager$^{12}$
Guillermo Torres$^4$, 
William F. Welsh$^{13}$
}
\affil{$^1$Fermilab Center for Particle Astrophysics, P.O. Box 500, Batavia, IL 60510}
\affil{$^2$Department of Astronomy and Physics, San Jose State University, San Jose, CA 95192}
\affil{$^3$NASA Ames Research Center, Moffett Field, CA 94035}
\affil{$^4$Harvard-Smithsonian Center for Astrophysics, 60 Garden St., Cambridge, MA 02138}
\affil{$^5$Niels Bohr Institute, Copenhagen University, DK-2100 Copenhagen, Denmark}
\affil{$^6$McDonald Observatory, The University of Texas, Austin, TX 78712-2059 USA}
\affil{$^7$Department of Astronomy, University of Florida, 211 Bryant Space Science Center, Gainesville, FL 32611-2055, USA}
\affil{$^8$Department of Astronomy and Astrophysics, Unvirsity of California, Santa Cruz, 95064}
\affil{$^{9}$National Optical Astronomy Observatory, Tucson, AZ 85719, USA}
\affil{$^{10}$Astronomy Department, UC Berkeley, Berkeley, CA 94720}
\affil{$^{11}$SETI Institute, 515 North Whisman Road, Mountain View, CA, 94043}
\affil{$^{12}$Department of Physics, Massachussets Institute of Technology}
\affil{$^{13}$San Diego State University, 5500 Campanile Drive, San Diego, CA 92182}

\begin{abstract}
We present and discuss five candidate exoplanetary systems identified with the \textit{Kepler} spacecraft.  These five systems show transits from multiple exoplanet candidates.  Should these objects prove to be planetary in nature, then these five systems open new opportunities for the field of exoplanets and provide new insights into the formation and dynamical evolution of planetary systems.  We discuss the methods used to identify multiple transiting objects from the \textit{Kepler} photometry as well as the false-positive rejection methods that have been applied to these data.  One system shows transits from three distinct objects while the remaining four systems show transits from two objects.  Three systems have planet candidates that are near mean motion commensurabilities---two near 2:1 and one just outside 5:2.  We discuss the implications that multitransiting systems have on the distribution of orbital inclinations in planetary systems, and hence their dynamical histories; as well as their likely masses and chemical compositions.  A Monte Carlo study indicates that, with additional data, most of these systems should exhibit detectable transit timing variations (TTV) due to gravitational interactions---though none are apparent in these data.  We also discuss new challenges that arise in TTV analyses due to the presence of more than two planets in a system.
\end{abstract}


\maketitle

\section{Introduction}

The discovery of dozens of transiting planets has enabled astronomers to characterize key physical properties of the planets, including their sizes, densities, atmospheric composition, thermal properties, and the projected inclination of the orbit with respect to the stellar spin axis \citep{char2007}.  Ground-based transit searches have surveyed many more stars than radial velocity planet searches, allowing them to discover relatively rare planets, such as giant planets with orbital periods of less than two days.  However, ground-based transit surveys are only efficient for large planets with relatively short orbital periods.   These strong detection biases and the likely dynamical instability of a system with multiple giant planets packed close to the host star, may explain why ground-based transit surveys have yet to detect a system with multiple transiting planets orbiting the same star.

The \textit{Kepler} mission was designed to detect terrestrial-size planets in the habitable zone of the host star, necessitating both a large sample size and sensitivity to a much larger range of orbital separations than ground-based surveys \citep{boru2010}.  The instrument is a differential photometer with a wide (105 square degrees) field-of-view (FOV) that continuously and simultaneously monitors the brightness of approximately 150,000 main-sequence stars.  A comprehensive discussion of the characteristics and on-orbit performance of the instrument and spacecraft is presented in \citet{koch2010}.

Its sensitivity to small planets over a wide range of separations gives \textit{Kepler} the capability of discovering multiple planet systems.  For closely packed planetary systems, nearly coplanar systems, or systems with a very fortuitous geometric alignment, \textit{Kepler} is likely to detect transits of multiple planets.  For systems with widely spaced planets or large relative inclinations, not all planets will transit, but some may still be detectable based on transit timing variations (TTVs) due to the gravitational perturbation of one or more non-transiting planets \citep{agol2005,holm2005}.  In other cases, non-transiting planets may be detectable by follow-up observations, such as radial velocity observations originally intended to measure the mass of the transiting planet(s) \citep{lege2009,quel2009}.

Radial velocity (RV) surveys have shown that giant planets often reside in multiple planet systems \citep{wrig2009}.  Given the large number of candidate planets identified by \textit{Kepler} \citep{boru2010}, it is expected that some fraction of them will be in multiple planet systems and a fraction of those will have multiple planets that transit.  In addition to the ability to characterize physical properites of each transiting planet, planetary systems with multiple transiting planets present several advantages.  For example, the fact that each planet formed from the same proto-planetary disk provides more powerful constraints for models of planet formation and orbital migration.  Moreover, these systems are quite powerful for studying the detailed orbital dynamics through transit timing variations (or lack thereof).  In some cases, the planet masses may be determined without measurements of stellar RV variations.  Systems with more than one transiting planet hold unique power in this regard as the period, orbital phase, and approximate size of the various planets are known.  This information can help significantly in finding a unique solution to an otherwise challenging and degenerate inversion problem~\citep{ford2007,nesv2008,mesc2010}.

In cases where RV measurements are able to measure the planet masses independently of TTVs, the two techniques can be combined to measure the mass and size of the host star without relying on stellar models \citep[see][]{agol2005,holm2005}.  In cases where RV observations are not practical (e.g., hot stars, fast rotators) or would require prohibitive observing time (i.e., faint stars), the detection of TTVs can be used to confirm that transit candidates orbit the same star---as opposed to being two objects transiting two stars blended within a single point-spread-function (PSF)---and to determine if the companions are of planetary mass.  Here multiple transiting systems are particularly powerful as the period, phase, and size of additional planets can be determined from the light curve.  This additional information can also help simplify the inverse problem~\citep{stef2007}.

We present five planetary candidate systems in which the transits of multiple objects can be seen in the first quarter of photometric data (a 33.5-day data segment from May 13 to June 15 UT, 2009) from the \textit{Kepler} spacecraft.  While not confirmed planet discoveries, these systems have passed several important tests that eliminate false-positive signals.  If all were ultimately shown to be planets, then these systems would contain four planets with radii smaller than three Earth radii (the smallest being two Earth radii), at least two pairs of planets in or very near a low-order mean-motion resonance (MMR), and one system with at least three distinct transiting planets.

For simplicity, we will refer to these objects as ``planets'' throughout this paper, recognizing that their confirmation as such is yet incomplete and that some of these transit signals may be due to other astrophysics.  The stellar references that we will use throughout this paper are \textit{Kepler} Objects of Interest (KOI) 152, 191, 209, 877, and 896 with the transiting planets denoted by ``.01'', ``.02'', etc. beginning in the order that they were identified with the transit detection software from the \textit{Kepler} pipeline.  Thus, the planet number designation does not necessarily reflect the order of the planets within each system.  We do not use letter designations, which by convention are reserved for confirmed planets.

This paper will proceed as follows.  First, we give the known properties of the host stars (\S \ref{secStar}).  In \S \ref{falsepositive} we discuss the photometric reduction and the algorithm used to identify the multiple candidates within each system.  We also outline the tests we have conducted to eliminate false-positive systems.  We present estimates of the orbital and physical properties of these objects should they prove to be planets (\S \ref{properties}).  In \S \ref{TTV} we discuss the possible future detection of transit timing variations based upon a Monte Carlo simulation of these candidate systems.  Finally, we discuss the implications of these results in \S\ref{secDiscuss}.

\section{Stellar properties\label{secStar}}

For each of the five stars, we obtained high resolution echelle spectroscopy from the McDonald Observatory 2.7m telescope Tull Coud\'e spectrometer with resolving power R=60,000.  We also obtained one spectrum of KOI 155 with the FIES spectrograph on the Nordic Optical Telescope and one spectrum of KOI 191, KOI 209, KOI 877, and KOI 896 with the Kitt Peak National Observatory 4-m telescope.  These spectra were obtained for the purpose of constraining effective temperature $T_{\text{eff}}$, surface gravity $\log g$, projected rotational velocity for the star $v\sin i$, and metallicity $[M/H]$.  The McDonald spectra were reduced and extracted using the IRAF echelle package \footnote{IRAF is distributed by the National Optical Astronomy Observatory, which is operated by the Association of Universities for Research in Astronomy (AURA) under cooperative agreement with the National Science Foundation.}.

In all cases, the spectroscopic analysis was done by matching the observed spectra to a library of synthetic spectra.  The synthetic spectra cover the wavelength region 5050-5360\,\AA \ (centered roughly on the Mg b lines).  The grid has coarseness of 250 K in $T_{\text{eff}}$, 1-4 km/s in $v_{\text{rot}} \simeq v \sin i$, and 0.5 dex in $\log g$ and $[M/H]$, implying uncertaintes of half those values.  For the host stars with low SNR spectra, we performed a diagnostic using the J$-$K color to verify that their compositions are consistent with solar metallicity, but the values we report are simply from template-matching with $[M/H]$ fixed at 0.  Table~\ref{starproperties} lists the resulting stellar parameters for all five stars as well as the instruments used in the observations.  All five stars apparently reside near or on the main sequence.

\begin{deluxetable}{crccccccccl}
\tablecolumns{11}
\tabletypesize{\scriptsize}
\tablewidth{0pc}
\tablecaption{Stellar properties and locations for the five candidate systems.\label{starproperties}}
\tablehead{
\colhead{KOI} & \colhead{KIC-ID} & \colhead{RA} & \colhead{DEC} & \colhead{Kepmag} & \colhead{$T_{\text{eff}}$} & \colhead{$\log g$} & \colhead{$[M/H]$} & \colhead{$v\sin i$} & \colhead{$v_{\text{abs}}$} & \colhead{$N_{\text{exp}}$}\\
\colhead{} & \colhead{} & \colhead{(2000)} & \colhead{(2000)} & \colhead{} & \colhead{($K$)} & \colhead{} & \colhead{} & \colhead{(km/s)} & \colhead{(km/s)} & \colhead{}
}
\startdata
152 & 8394721 & 20 02 04.1 & 44 22 53.7 & 13.9 & 6500 & 4.5  & 0 (fixed) & 14  & -22.32 & $2^{a,b}$ \\
191 & 5972334 & 19 41 08.9 & 41 13 19.1 & 15.0 & 5500 & 4.5  & 0 (fixed) & 0   & -62.98 & $1^b$ \\
209 & 10723750 & 19 15 10.3 & 48 02 24.8 & 14.2 & 6100 & 4.1  & -0.05     & 7.8 & -12.78 & $4^b$ \\
877 & 7287995 & 19 34 32.9 & 42 49 29.9 & 15.0 & 4500 & 4.0  & 0 (fixed) & 0   &  0.531 & $2^{b,c}$ \\
896 & 7825899 & 19 32 14.7 & 43 34 52.9 & 15.3 & 5000 & 4.0  & 0 (fixed) & 1   & -21.28 & $1^c$ \\
\enddata
\tablecomments{
Each system was analyzed by matching the observed spectra to the CfA library.\\
Telescopes used for these observations:\\
$a$, Nordic Optical Telescope, FIES\\
$b$, McDonald Observatory 2.7m, Tull Coud\'e\\
$c$, Keck-HIRES
}
\end{deluxetable}

\section{\textit{Kepler} data and photometric analysis\label{falsepositive}}

\subsection{Transit identification}

Each of these systems was found using the Transiting Planet Search Pipeline (TPS) which identifies significant transit-like features, or Threshold Crossing Events (TCE), in the \textit{Kepler} light curves \citep{jenk2010}.  Data showing TCEs are then passed to the Data Validation (DV) pipeline \citep{wu2010}.  The purpose of the DV pipeline is twofold: 1) to fit a transiting planet model to the data, remove it from the light curve, and to return the result to TPS in an effort to find additional transit features and 2) to complete a suite of statistical tests that are applied to the data after all TCEs are identified in an effort to assess the likelihood of false-positives.  The binary discrimination statistics and the motion detection statistic, in particular, speak to the likelihood of astrophysical false positives such as grazing eclipses and diluted eclipsing binaries.  These statistics are described below.

After pipeline data processing and the photometry extraction, the time series is detrended with a running 1-day median.  All observations that occur during transit are not included in the evaluation of the median.  The transit lightcurve is modeled using the analytic expressions of \citet{mand2002} using the non-linear limb darkening parameters that \citet{clar2000} derived for the \textit{Kepler} bandpass.  A first estimate for $M_*$ and $R_*$ is obtained by comparing the derived stellar $T_{\text{eff}}$ and $\log g$ values, obtained using the method of \citep{vale1996}, to a set of CESAM \citep{more1997} stellar evolution models computed in steps of $0.1\,M_{\odot}$ for solar composition.  We note that in some cases, particularly KOI 877 which shows significant spot modulation, there is some residual noise power at time scales longer than the transit duration.  While this noise causes small baseline fluctuations in the out-of-transit portion of the light curves, it does not significantly affect estimates of the transit model parameters (e.g., the transit depth).

\begin{figure}
\includegraphics[width=0.45\textwidth]{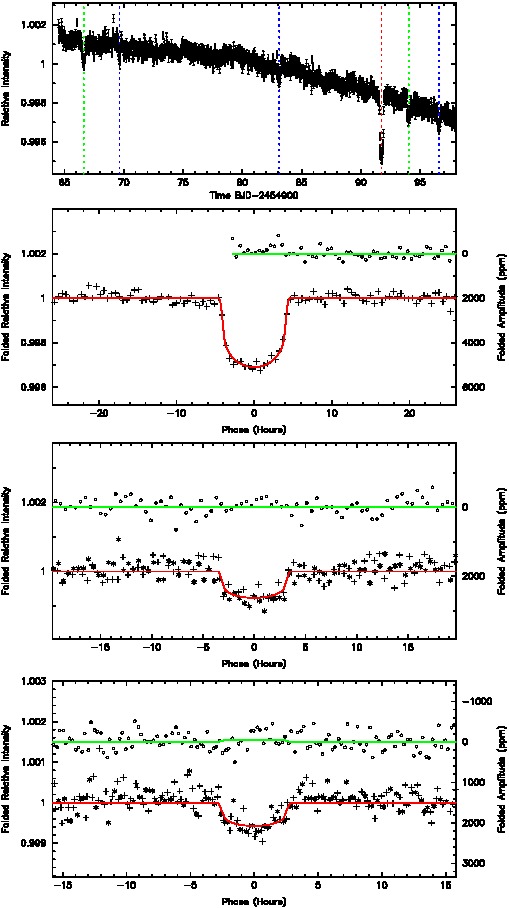}
\caption{Unbinned light curve and transit models for the candidates KOI 152.01 (top), 152.02 (middle), and 152.03 (bottom).  The lower curves in each panel show the transit while the vertically-offset, upper curves show the data at phase $= 0.5$.  The ``+'' and ``*'' symbols are for even and odd transits.  The top-most panel shows the raw lightcurve with the transits for the three candidates marked with vertical lines.
\label{lights152}}
\end{figure}

\begin{figure}
\includegraphics[width=0.45\textwidth]{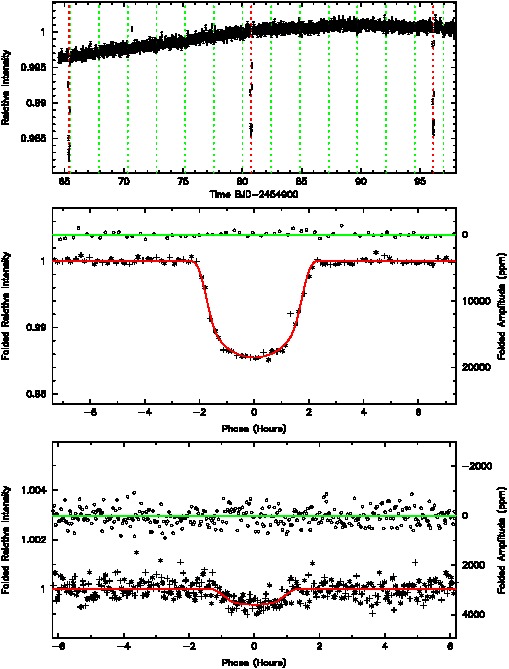}
\caption{Light curve and transit models for the candidates KOI 191.01 (top) and 101.02 (bottom).
\label{lights191}}
\end{figure}

\begin{figure}
\includegraphics[width=0.45\textwidth]{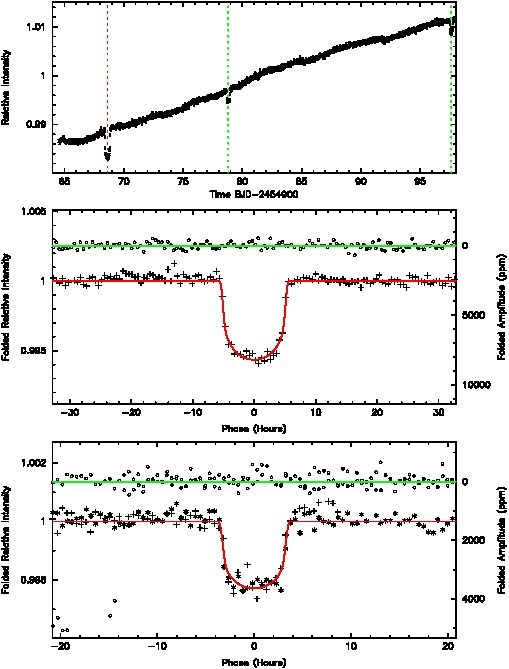}
\caption{Light curve and transit models for the candidates KOI 209.01 (top) and 209.02 (bottom).
\label{lights209}}
\end{figure}

\begin{figure}
\includegraphics[width=0.45\textwidth]{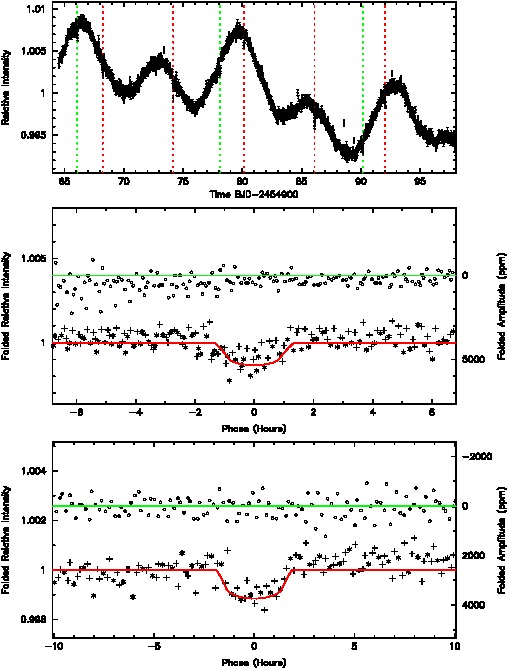}
\caption{Light curve and transit models for the candidates KOI 877.01 (top) and 877.02 (bottom).
\label{lights877}}
\end{figure}

\begin{figure}
\includegraphics[width=0.45\textwidth]{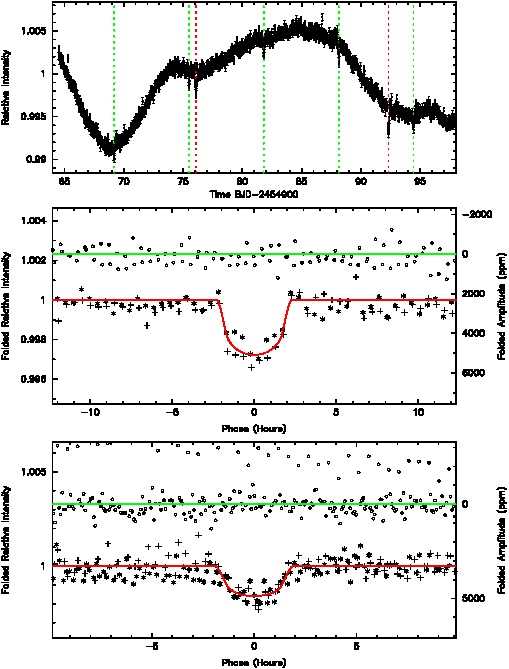}
\caption{Light curve and transit models for the candidates KOI 896.01 (top) and 896.02 (bottom).
\label{lights896}}
\end{figure}

With $M_*$ and $R_*$ fixed to their initial values, a transit fit is then computed to determine the orbital inclination, planetary radius, and depth of the occultation (passing behind the star) assuming a circular orbit.  The best fitting model is found using a Levenberg-Marquardt minimization algorithm \citep{pres1992}.  This model is then removed from the lightcurve and the residuals are used to characterize the next transit candidate identified by TPS.  The light curves and transit models for these five systems are shown in Figures~\ref{lights152}, \ref{lights191}, \ref{lights209}, \ref{lights877}, and \ref{lights896}.  The significance of individual transit events for each planet and the significance of the folded transit event (in terms of canonical ``sigmas'') are given in the first two columns of Table \ref{FPtab}.  Note that these statistics come from the pipeline processing and are not calculated for systems where there is only one transit of the most significant transit event (KOIs 152 and 209); the additional candidates in these two systems were found manually.

\subsection{Candidate vetting from \textit{Kepler} data}

The depths of planetary transits should be consistent from orbit to orbit as well as evenly spaced in time.  Eclipsing stellar binaries, on the other hand, generally have primary and secondary eclipses with different depths that are frequently spaced asymmetrically in time.  The binary discrimination test includes two metrics: the Odd/Even statistic and the Epoch statistic.  The Odd/Even statistic is a comparison between the depth of the phase-folded, odd-numbered transits and the depth of the phase-folded, even-numbered transits.  The Epoch statistic compares the timing of the odd and even-numbered transits.  Both statistics are constructed as $\chi^2$ distributions, and the significance (reported in Table~\ref{FPtab}) of the statistic is obtained by evaluating the $\chi^2$ cumulative distribution function for the appropriate number of statistical tests.  The significance is the probability that the statistic is consistent with the binary interpretation.

The motion detection statistic identifies objects with flux-weighted centroids that are highly correlated with a transit signature derived from the flux timeseries.  The statistic is a $\chi^2$ variable with two degrees of freedom (row and column).  The probability of producing a statistic of equal or lesser value is reported as ``Motion'' in Table~\ref{FPtab}.  It is computed by evaluating the $\chi^2$ cumulative distribution function at the value of the detection statistic given two degrees of freedom.  The complement of this value is reported so that values near unity represent a small likelihood of a correlation.  Centroids that are highly correlated with the transit signature are indicative of a crowded photometric aperture---a warning that the transit could be due to a nearby eclipsing binary diluting the flux of the target star.  Such cases are denoted by a motion significance near zero.  A full description of DV statistics is given in \citet{wu2010}.

Outside of the pipeline, all TCEs are fitted with a planet transit model as described in \citet{bata2010}.  Those yielding an estimated planet radius less than 2 $R_J$ are assigned a \textit{Kepler} Object of Interest (KOI) number.  The modeling returns an independent test of the Odd/Even statistic (Depth-sig in Table~\ref{FPtab}), expressed in units of the standard deviation.  The modeling also tests for the presence of secondary eclipses (or occultations) at phase $= 0.5$ and reports this as Eclipse-sig in Table~\ref{FPtab}, also in units of the standard deviation.  We note that the transit modeling "Depth-sig" statistic and the DV "Odd/Even" statistic are not identical as they arise from different analyses (especially when the odd and even transit depths are not significantly different) but will be approximately equivalent when a significant detection is seen.

\begin{deluxetable}{rccccccc}
\tablecolumns{6}
\tablewidth{0pc}
\tablecaption{Planetary candidate transit detection and validation statistics.\label{FPtab}}
\tablehead{
\colhead{KOI} & \colhead{Transit-Sig} & \colhead{Transit-Sig} & \colhead{Odd/Even} & \colhead{Epoch} & \colhead{Motion} & \colhead{Depth-Sig} & \colhead{Eclipse-Sig} \\
\colhead{} & \colhead{(Single)} & \colhead{(Folded)} & \colhead{} & \colhead{} & \colhead{} & \colhead{} & \colhead{}
}
\startdata
152.03 & -- & -- &  -- & -- & --  & 0.6 & -0.5  \\
    02 & -- & -- &  -- & -- & --  & 1.3 & -1.0  \\
    01 & -- & -- &  -- & -- & --  &  -- &  --   \\
191.02 & 4.2&13.0& 0.71&0.95&0.56 & 0.0 & -0.5  \\
    01 &104 &211 & 0.43&0.97&0.02 & 1.8 & -2.6  \\
209.02 & -- & -- &  -- & -- & --  & 0.3 & -0.7  \\
    01 & -- & -- &  -- & -- & --  &  -- &  --   \\
877.01 & 4.9&13.0& 0.14&0.97&0.93 & 1.7 &  2.5  \\
    02 & 5.0&10.4& 0.83&0.97&0.94 & 0.5 &  0.5  \\
896.02 & 5.3&13.3& 0.11&0.96&0.61 & 0.0 & -4.0  \\
    01 & 7.9&15.7& 0.16& -- &0.45 & 1.9 &  1.5  \\
\enddata
\end{deluxetable}

\begin{figure}
\includegraphics[width=0.225\textwidth]{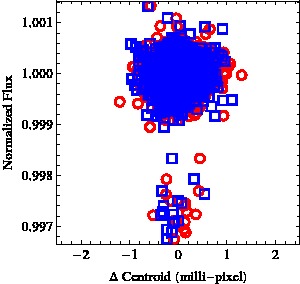}
\includegraphics[width=0.225\textwidth]{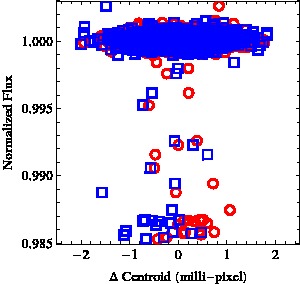}
\includegraphics[width=0.225\textwidth]{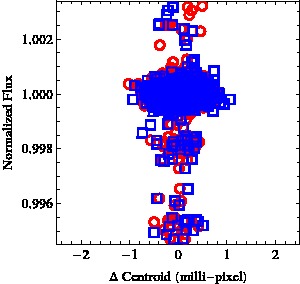}
\includegraphics[width=0.225\textwidth]{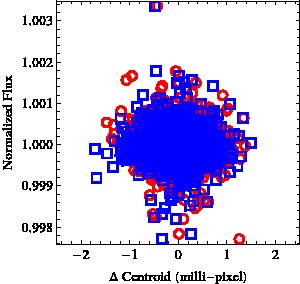}
\includegraphics[width=0.225\textwidth]{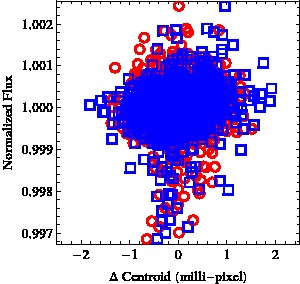}
\includegraphics[width=0.225\textwidth]{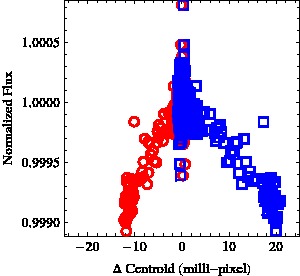}
\caption{Relative flux vs. centroid position for KOIs 152, 191, 209, 877, and 896 beginning in the top left corner.  The plot in the bottom right is from a background eclipsing binary and is presented to illustrate the difference in the data between good planetary candidates and background eclipsing binaries.  The blue squares correspond to rows while the red circles are columns on the CCD.
\label{rainplots}}
\end{figure}

Figure~\ref{rainplots} shows the normalized relative flux vs. centroid position (referred to as a ``rain plot'') as described in \citet{bata2010} for each target star.  Here, the relative flux is plotted against the relative centroid position along rows and columns.  A centroid shift that is highly correlated with the transit signature would appear as a diagonal deviation in the plot (see bottom-right panel of Figure~\ref{rainplots}) whereas uncorrelated centroids ``rain down'' vertically under the out-of-transit points.  No significant correlations can be readily identified in the rain plots for any of the candidates presented here except KOI 191.  This is reflected in the Motion statistics reported in Table~\ref{FPtab}.  KOI 191 has a correlation with a 98\% significance (taking the complement of the value reported in Table~\ref{FPtab})---comparable to a 2.3-sigma detection.  A correlation does not necessarily rule out the planetary interpretation, rather it should be interpreted as a warning that the photometric aperture is crowded.  Additional analysis or observational follow-up is required to determine the location and magnitude of each star in the vicinity and, ultimately, the origin of the transit-like features.

None of the DV binary discrimination statistics reported in Table~\ref{FPtab} are significant at the 3-sigma level or higher (corresponding to a significance of 0.997 or larger).  However, DV statistics for KOI 152 and 209 are not available due to the fact that the dominant transit feature in the Quarter 1 light curve appears only once.  A planet transit model cannot be fitted to a light curve with a single transit with DV.  Consequently, it does not get filtered and passed back to TPS for the detection of the shallower, shorter-period transits.  The shorter-period events are subjected to light curve modeling and the associated statistics are reported.

\subsection{Candidate vetting from ground-based data}

\begin{figure}
\includegraphics[width=0.225\textwidth]{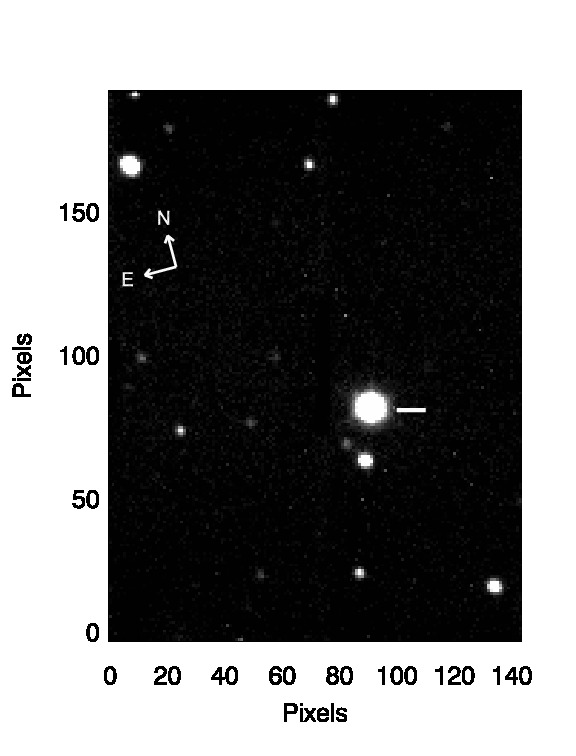}
\includegraphics[width=0.225\textwidth]{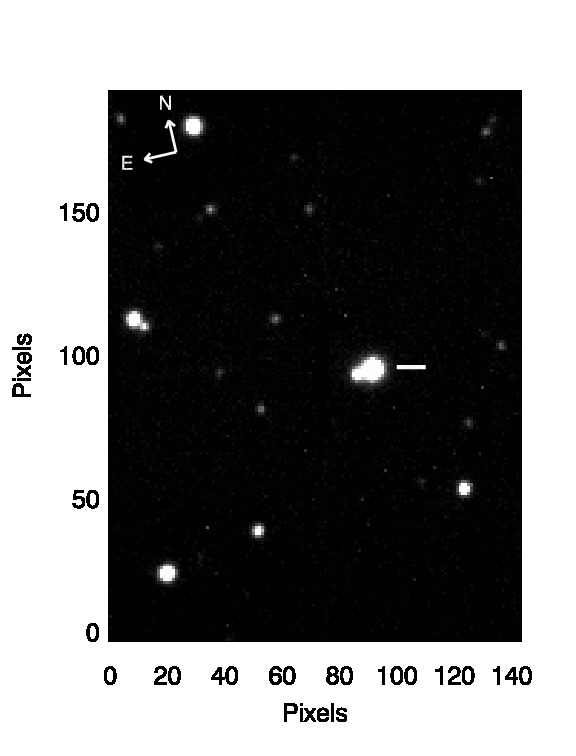}
\includegraphics[width=0.225\textwidth]{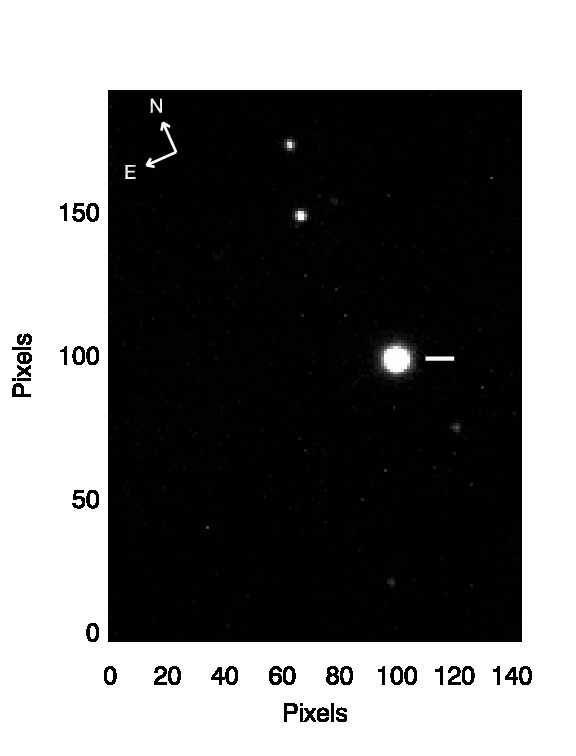}
\includegraphics[width=0.225\textwidth]{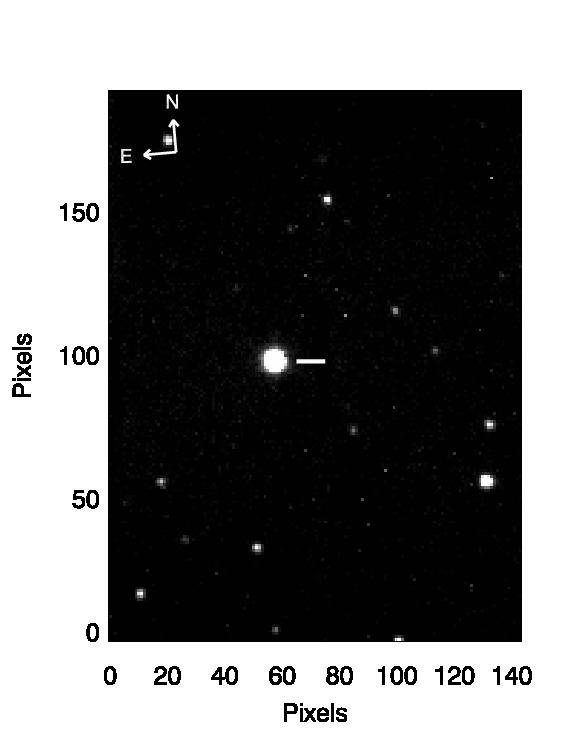}
\includegraphics[width=0.225\textwidth]{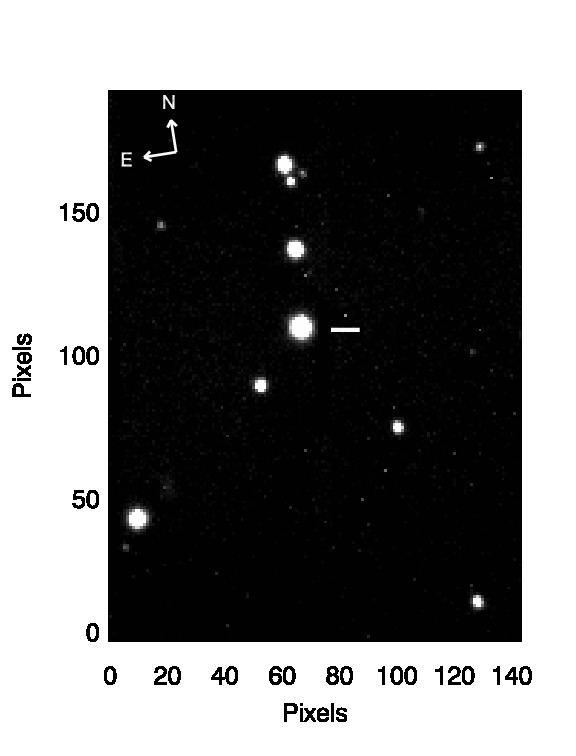}
\caption{Optical images of all five \textit{Kepler} stars taken in seeing of 0.7-0.9 arcsec and clear skies by the guider camera on the Keck-HIRES instrument.  The images reveal neighboring stars that may affect the interpretation of the transits, notably their depths and the existence of eclipsing binaries.  In order from the top left the images are of KOIs 152, 191, 209, 877, and 896.
\label{keckimages}}
\end{figure}

In order to identify neighboring stars that are located within a few arcsec of the target star, we obtained optical images of all five stars using the photometrics CCD guide camera on the HIRES spectrometer on the Keck 1 telescope (shown in Figure~\ref{keckimages}).  The images were obtained with no filter over the CCD that has high sensitivity from 400-800 nm, mimicking the CCD on the \textit{Kepler} camera.  The images were taken in seeing of 0.7 - 0.9 arcsec and clear skies.  Each image is $43 \times 57$ arcsec, with 0.30 arcsec/pixel.  Such neighboring stars bring two concerns, diluting the transit depths and possibly being eclipsing binaries that cause the transit signal.  In the latter case, one or more of the candidate planets may be false positives.

For KOI 152, the Keck image shows two neighboring stars located south and east of the KOI (which is the brightest star in the field).  The brighter of the two neighbors is 5.4 arcsec SE and 3.8 mag fainter.  The fainter neighboring star is 4.5 arcsec to the southeast and 6.6 mag fainter.  The Keck image of KOI 209 reveals no neighboring stars down to 20th mag.  The Keck  image of KOI 896 reveals one neighboring star 7.3 arcsec to the southeast that is 2.9 mag fainter, and another neighboring star 8.1 arcsec to the north that is 1.5 mag fainter.  The wings of both neighboring stars encroach into the \textit{Kepler} photometric aperture.  The Keck image of KOI 877 shows no neighboring stars down to 20th mag.  We have not examined these neighboring stars to determine if they are eclipsing binaries.  However, the centroid statistics given above indicate that, should they prove eclipsing binaries, they are unlikely to be the cause of the observed transit signature.

For KOI 191, the Keck image reveals a neighboring star located 1.5 arcsec east of the main star and 2.6 mag fainter.  As stated above, KOI 191 exhibits some correlation in the rain plots which indicates a crowded field.  The light curve from KOI 191 also shows an additional, periodic transit feature.  The ephemeris of this feature is $T_{\text{c}} - 2454900 BJD) = 65.6589 + E \times 0.7086$ Days.  Its V-shaped transit shape (0.2 mmag depth) and its two-hour duration indicate that a faint eclipsing binary is also blended with the target.

\subsection{Blender analysis}

False positive scenarios were investigated for the five systems by exploring the possibility that the \textit{Kepler} photometry is the result of contamination of the light of the candidate by an eclipsing binary along the same line of sight, a ``blend''.  Given that the centroid motion statistics discussed above rule out a large fraction of the background contaminants, we focused here on hierarchical triple systems in which the candidate and the binary are at the same distance.  Angular separations in these cases would usually be too small to generate significant centroid motion.

We modeled the \textit{Kepler} photometry of each candidate assuming it is the result of the brightness variations of an eclipsing binary being attenuated by the (typically) brighter candidate star.  For KOIs with two or more signals in the light curve, we modeled the light curves at each period separately and accounted for possible blends at the other period(s) by incorporating extra dilution consistent with those other stars.  In these cases, whether the blended eclipsing systems at each period are related or not (i.e., in a hierarchical quadruple system, for candidates with two signals) is immaterial for the purposes of modeling the light curves.  The objects composing the binary are referred to as the ``secondary'' and ``tertiary'', and the candidate is the ``primary''.  The procedure closely follows that described by \citet{torr2004} and consists of calculating synthetic light curves that result from the three objects for a wide range of eclipsing binary parameters, and comparing those light curves
against the \textit{Kepler} photometry for the candidate, in a $\chi^2$ sense.  We regard as acceptable any blend scenario that results in a synthetic light curve giving a $\chi^2$ for the fit that is not significantly different (at the 3-$\sigma$ confidence level) from a planet model fit.

The brightness variations of the binary are generated using detailed calculations including limb darkening, gravity brightening, reflection, and proximity effects.  The properties of the candidate are tightly constrained by the spectroscopic parameters in Table~\ref{starproperties} and were held fixed.  The parameters of the binary components were taken from model isochrones by \citet{gira2000}, parametrized in terms of their mass.  The secondary and tertiary masses were allowed to vary over wide ranges (0.1--1.4~$M_{\sun}$) in order to fit the light curve, and the inclination angle was also a free parameter.  By comparing the quality of the light-curve fits over these ranges, we are able to constrain the properties of the secondaries and tertiaries that provide acceptable fits, and we can reject other blends.  We consider hierarchical configurations of two types: ones in which the tertiary is a star, and ones in which it is a planet contributing no light.  In the latter case the size of the planet is a free parameter, which we varied between 0.1 and 2.0~$R_{\rm Jup}$.  We account for the additional stars identified by high resolution imaging by including the proper amount of extra light in our models.

In all five cases we find that configurations with stellar tertiaries are inconsistent with the \textit{Kepler} light curves, for any size secondary.  Thus, hierarchical triples involving a stellar binary are ruled out.  When the tertiary is allowed to be a planet, we find that there is a range of possible solutions that yield acceptable matches to the \textit{Kepler} photometry, often times as good as obtained from a single star and planet model.  However, many of those solutions can also be excluded on other grounds as described below.

For KOI 152, which is a mid F dwarf, the only configurations consistent with the \textit{Kepler} light curves involve secondary stars that are at least as massive as the primary---and therefore almost as bright or brighter.  In all cases the planetary companions are roughly 0.5~$R_{\rm Jup}$ in size.  Such bright secondaries are unlikely as they would have been seen spectroscopically, unless the velocity difference with the target star happens to be very small so that the lines are indistinguishable.  This cannot be completely ruled out if the two stars are in a wide orbit around each other, but it would require a special set of circumstances given the constraints on the centroid motion.  Thus, except for this particular case of a bright secondary, our analysis supports the planetary interpretation.

For KOI 209, a mid to late F dwarf, we again find that the only blend configurations that fit the \textit{Kepler} light curve involve a secondary that is as bright as the primary or brighter, and a tertiary about the size of Jupiter.  A planetary interpretation is again favored.

Our blend simulations for KOI 896, an early K star, again indicate that the secondaries required for a good fit to the light curve are similar in brightness to the primary, and the tertiaries have sizes between 0.4 and 0.5~$R_{\rm Jup}$. Solutions with smaller secondary stars are visibly worse.  Therefore, except for the somewhat artificial case of twin stars, the \textit{Kepler} photometry favors a planetary interpretation.

For KOI 191.02 the light-curve fits allow blend scenarios in which the secondaries are smaller than the late G-type primary, down to a spectral type of late K.  The tertiaries tend to have radii around 0.3~$R_{\rm Jup}$.  In the case of KOI 191.01 we find that the light curve prefers secondaries that are as bright as or brighter than the primary and tertiaries of $\sim$1.5~$R_{\rm Jup}$.  Both of these results take account of the fact that there is extra light from the 1.5 arcsec companion described earlier, which we have considered here to be an unrelated background eclipsing binary.

KOI 877 is a late K dwarf.  In addition to the blend solutions with bright secondary stars, we find acceptable fits to the light curves with secondaries up to two magnitudes fainter than the primary (spectral type M2-M3), which might not be noticed spectroscopically.  The size of the tertiaries in these cases is about 0.4~$R_{\rm Jup}$.  To rule out such configurations, additional observational constraints are needed, such as accurate multi-band photometry out of transit to check for color inconsistencies \citep[e.g.,][]{odon2006}.

We note that two eclipsing binaries diluting each other, perhaps in a hierarchical quadruple system \citep[e.g.,][]{pile2007a} is among the most likely, still-viable false positive scenarios that involves no planets.  To our knowledge, there are no well-established cases of stellar systems in which more than one body passes in front of the primary.  
The All-Sky Automated Survey \citep{pile2007a} identified seven systems that were termed ``double eclipsing binaries'': physically bound quadruples or blended independent binaries in which each pair of binaries eclipses.  Such systems are viable false positive scenarios since they produce no astrometric centroid motion, can be dynamically stable, and are astrophysically reasonable.  (Two small stars orbiting the same primary can be dismissed---except perhaps for KOI 191---because at the periods observed here, such systems are not stable even on very short timescales.)  The probability that all of the transiting planet candidates in a multiple system are produced by eclipsing binary false positives is less than $\sim 10^{-3}$ since roughly this fraction of the \textit{Kepler} stars are eclipsing binaries \citep{prsa2010} and the two must be blended into a single PSF.  Similar arguments, using the approximate number of \textit{Kepler} planetary candidates \citep{boru2010b}, show that the scenario of two independent stars with independently orbiting planets must be smaller still.  

In summary, the above blend analyses suggest that KOI 152, 209 and 896 are very likely true planets orbiting a common star.  KOI 191 and 877, on the other hand, could be blends of two planets orbiting two separate stars.  In none of these systems was a configuration of transiting stars preferred over transiting planet-sized objects.  

\section{Planetary and system properties\label{properties}}

\subsection{Orbital properties}

Of the five candidate planetary systems, there are four pairs of planets with a well-characterized orbital period ratio.  Two pairs are near the 2:1 MMR (KOI 152.02/01 and KOI 877.01/02), one pair is slighly outside the 5:2 commensurability (KOI 896.02/01), and one pair is hierarchical with a period ratio exceeding 6:1 (KOI 191.02/01).  The proximity of KOI 152 and 877 to the 2:1 MMR hints that these systems are likely to have large TTVs.  However, the timescale for large TTVs in resonant systems can be quite long.  For a system of Neptune-mass planets librating about a 2:1 MMR, we would not expect to detect TTVs based on the Q0 and Q1 data presented.  Moreover, for librating systems with smaller masses or that are very close to exact resonance, the time needed to distinguish a TTV signal from a constant period lengthens, also requiring additional data.

Inferring the relative frequency of planets with various orbital spacings will require a population analysis that corrects for the geometric transit probabilities.  Of course, there may be additional non-transiting planets, or small transiting planets, between KOI 191.02 and 191.01 or in any of the candidate systems presented.  In some cases, TTVs or radial velocity follow-up could identify such planets.  In other cases, the non-transiting planets will remain undetected, further complicating a population analysis.  The sample of multiple planet candidate systems presented here is too small---and not necessarily unbiased---for such an analysis.

For two of the transit candidates presented, only one transit has been observed.  There is a small chance that a second transit occured during a data gap.  However, this is {\em a priori} unlikely and the extended transit durations also support large orbital periods.  The periods listed in Table~\ref{tabPeriods} are lower limits based upon the non-observation of a second transit in the data.  Also included, however, are estimates based upon the transit duration assuming a circular orbit and a central transit of the planet.  With these latter estimates, it is tempting to identify the KOI 209 system as lying near a 3:1 MMR and the KOI 152 system as being near a 4:2:1 MMR.  However, we caution that the uncertainty in the orbital period estimated from a single transit is far too large to have confidence that these systems are near MMR.  Nonetheless, \citet{yee2008} show that orbital period estimates based upon single transit durations need not be as conservative as the estimates that we state.

For each neighboring pair of planet candidates, we measure the ratio $\xi \equiv (D_{in}/D_{out})(P_{out}/P_{in})^{1/3}$, where $D$ is the transit duration and $P$ is the orbital period.  In each case the ratio is near unity (see Table~\ref{orbittable}), as expected for a pair of objects on circular and coplanar orbits around a common star.  We compare these ratios to the results of Monte Carlo simulations of the distribution of $\xi$ (denoted $\xi_{\text{MC}}$) for pairs of planets on circular and coplanar orbits with orbital periods similar to those observed.  We assume random viewing angles, subject to the constraint that both planets transit.  The 16th, 50th, and 84th percentile of these distributions are included in Table~\ref{orbittable}.  In no case do we find evidence for a large eccentricity or for a blend of multiple stars each with one transiting object.  The largest deviation from unity is that for KOI 191, but in this case the observed and expected values are very close, due to the large ratio of semimajor axes.  In the cases of 209 and 896, the ratio is slightly less than expected for coplanar, circular orbits.  However, this could be easily reconciled if one planet in each system were to have a modest eccentricity ($\sim 0.05$ for KOI 209, $\sim 0.1$ for KOI 896).

\begin{deluxetable}{rrrcccc}
\tablecolumns{7}
\label{tabPeriods}
\tablewidth{0pc}
\tablecaption{Orbital periods and transit epochs for the candidate planets.\label{orbittable}}
\tablehead{
\colhead{Candidate} & \colhead{Transit Epoch} & \colhead{Period} & \colhead{Period Ratio} &\colhead{Transit Duration} & \colhead{$\xi$} & \colhead{$\xi_{\text{MC}}$}\\
\colhead{} & \colhead{BJD $-2454900$} & \colhead{(Days)} & \colhead{(vs. inner)} & \colhead{(Days)} & \colhead{(obs.)} & \colhead{(predicted)}
}
\startdata
152.03 & $69.622 \pm 0.0053$    & $13.478 \pm 0.0098$ & -      & $0.2071 \pm 0.0022$ & --     & --     \\
02     & $66.630 \pm 0.0079$    & $27.406 \pm 0.0150$ & 2.03   & $0.2823 \pm 0.0060$ & 0.9291 & $1.10^{+0.46}_{-0.09}$ \\
01     & $91.747 \pm 0.0026$    & $>27 \quad (51.9)$  & (3.85) & $0.3432 \pm 0.0013$ & 1.0188 & $1.08^{+0.36}_{-0.07}$ \\ \hline
191.02 & $65.50  \pm 0.16$      & $2.420 \pm 0.0006$  & -      & $0.0948 \pm 0.0016$ & --     & --     \\
01&$65.3847 \pm 4\times 10^{-4}$&$15.359 \pm 0.0004$  & 6.347  & $0.1494 \pm 0.0002$ & 1.1751 & $1.15^{+0.60}_{-0.13}$ \\ \hline
209.02 & $78.822  \pm 0.0046$   & $18.801 \pm 0.0087$ & -      & $0.2884 \pm 0.0018$ & --     & --     \\
01     & $68.635  \pm 0.0036$   & $>29 \quad (49.3)$  & (2.62) & $0.4252 \pm 0.0007$ & 0.9429 & $1.12^{+0.68}_{-0.11}$ \\ \hline
877.01 & $103.952 \pm 0.0028$   & $5.952 \pm 0.0024$  & -      & $0.0962 \pm 0.0012$ & --     & --     \\
02     & $114.227 \pm 0.0051$   & $12.039 \pm 0.0077$ & 2.023  & $0.1192 \pm 0.0021$ & 1.0204 & $1.08^{+0.47}_{-0.07}$ \\ \hline
896.02 & $107.051 \pm 0.0028$   & $6.311 \pm 0.0024$  & -      & $0.1278 \pm 0.0016$ & --     & --     \\
01     & $108.568 \pm 0.0024$   & $16.242 \pm 0.0075$ & 2.574  & $0.1916 \pm 0.0017$ & 0.9144 & $1.11^{+0.55}_{-0.10}$ \\
\enddata
\tablecomments{
The periods and period ratios listed in parentheses are estimates based upon the duration of a single transit.  The values listed for the predicted estimates $\xi_{\text{MC}}$ correspond to the median with the errors corresponding to the 16th and 84th percentiles.
}
\end{deluxetable}

\subsection{System coplanarity}

One important question surrounding these multi-candidate systems is the geometric probability that \textit{Kepler} would see both planets transiting.  Following \citet{rago2010}, and based on the method of \citet{boru1984}, we calculate this probability by considering the area of the region on the celestial sphere, centered on the star, that is aligned to see both planets transit \citep[see also][]{beat2010,gill2010}.  We also note here the analytical approximation given by \citet{rago2010} for the probability of observing both planets transit as a function of the true mutual inclination between the planets, $\phi$.  The result is different in low and high mutual inclination regimes, with the critical angle between the regimes as $\phi_{\rm crit} \simeq R_*/a_1$ where $R_*$ is the radius of the star and $a_1$ is the orbital distance of the innermost, transiting planet.  In the low mutual inclination regime, the probability of seeing both planets transit is $R_*/a_2$ where $a_2$ is the orbital distance of the outer planet.  Thus, the probability of seeing both planets transit is equal to the probability that the more distant planet transits.

In the high mutual inclination regime, this is no longer true, and only observations along the line of nodes of the orbital planes will see both planets transit \citep{koch1995,holm2005}.  In this regime, the probability of observing a transit is $R_*^2/(a_1a_2 \sin \phi)$, significantly lower than the probability in the low inclination regime.  With the three candidates seen in the KOI 152 system, the probability is more complicated as another mutual inclination angle and mutual nodal angle are required to specify the system.  Unlike in the two-planet case, it is easy to construct high mutual inclination systems where no observer would see three-planets transit.  If both mutual inclinations are low (i.e., below $1.5^{\circ}$), then the probability is $P \simeq R_*/a_3 \simeq 0.017$, where $a_3$ is the orbital distance of the third planet.  Introducing a larger mutual inclination between any pair of planets can significantly reduce this probability.  Even if one pair of planets has a mutual inclination of only 10$^{\circ}$, the probability of seeing all three transit drops to $\sim 0.0025$.  Based upon these probabilities, if these three objects are confirmed to be multiple planetary systems, then they are very likely coplanar to within a few degrees.

\begin{figure}
\includegraphics[width=0.35\textwidth,angle=90]{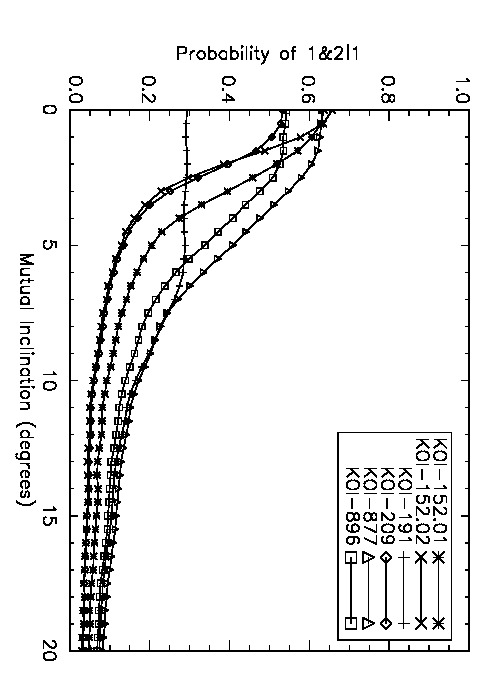}
\caption{Geometric transit probabilities, assuming that the inner planet transits, as a function of mutual inclination for multi-transiting systems using the method described in the text and discussed in \citet{rago2010}.  The curves shown are for the 4 double-transiting systems, the KOI 152.02/01 pair, and the KOI 152.03/02 pair.  These calculations assume solar-like stars and neglect the size of the planet.
\label{probmulti}
}
\end{figure}

Next, we consider the expected number of similar systems for which the outer planet does not transit.  This requires a calculation of the probability of seeing the outer planet transit given that the inner planet is known to transit.  This can also be answered with the model of \citet{rago2010} and also depends strongly on the mutual inclination.  Instead of providing analytical estimates, Figure~\ref{probmulti} shows a Monte Carlo numerical calculation of the fraction of random observers that see both planets transit out of those observers who see the inner planet transit.  All of the KOI systems are shown in this figure, including two curves for KOIs 152.02/152.01 and 152.03/152.02.

Even in the coplanar case, for each observed multi-transiting system there are $a_2/a_1$ multiple systems where only the inner planet transits.  Hierarchical systems like KOI 191, must have at least 3.5 times as many counterparts where only the 2-day period planet is seen in transit.  If any of these systems have large mutual inclinations, the number of implied similar systems increases considerably.  The short orbital periods and the likely near-coplanar state of these systems has implications for their formation.  If these planets formed beyond the snow line, some mechanism must be invoked to bring them planets to their current locations.  We note that none of these systems are candidates for formation by Kozai cycles with tidal friction due to perturbations by a distant stellar companion \citep{fabr2007}, since the presence of the other planet would shut off Kozai effects.

The two major classes of remaining theories for moving these planets in are planet-planet scattering \citep{rasi1996} and disk migration \citep{gold1980,lin1996}.  Planet scattering tends to excite orbital eccentricity and inclination and has difficulty migrating planets into short period orbits.  Disk migration is able to migrate multiple planets into short period orbits and tends to damp inclination.  The near-resonant ratios of the observed systems favors the disk migration hypothesis.  However, continued migration after resonant trapping excites eccentricities and inclinations \citep{lee2002,lee2009}.

\subsection{Physical properties\label{physparams}}

We consider the physical properties of planets (e.g., mass and composition) that can be derived from the \textit{Kepler} photometry.  The primary motivation for this effort is to identify plausible masses for these planets in order to conduct a Monte Carlo study of TTV signals (Section~\ref{TTV}).  For a given radius we estimate a range of masses that depends heavily on the possible bulk compositions.  For planets with radii larger than Saturn, the planet mass is largely indeterminate because of the transition in the mass-radius relation from a Coulomb to an electron degeneracy dominated equation of state.  The mass-radius curve turns over, so an object with a $1 R_J$ radius could be anything between a sub-Saturn mass planet (e.g., HAT-P-12b, $0.2 M_J$, \citet{hart2009}) to a brown dwarf (e.g., CoRoT-3b, $21.7 M_J$, \citet{dele2008}).  For a planet radius up to that of Neptune, the planet mass is constrained better between low-density objects rich in gas and volatiles and the rocky, iron-rich and high-density super-Earths.  Interpreting the bulk composition of such planets is more difficult due to the degeneracies that arise with materials having different equations of state.  The measured planetary radius and the expected mass ranges of the candidate planets is shown in Table~\ref{planetproperties}.

\begin{deluxetable}{rrrcc}
\tablecolumns{5}
\tablewidth{0pc}
\tablecaption{Planetary radii, fractional error, and likely range of masses.\label{planetproperties}}
\tablehead{
\colhead{Candidate} & \multicolumn{2}{c}{Planet Radius} & \colhead{Fractional Radius} & \colhead{Mass Range}\\
\colhead{} & \colhead{($R_J$)} & \colhead{($R_E$)} & \colhead{Error} & \colhead{}
}
\startdata
152.03 & 0.30 & 3.36 & 7.8\% & $9-30\ M_E$   \\
02     & 0.31 & 3.47 & 6.7\% & $9-30\ M_E$   \\
01     & 0.58 & 6.50 & 1.2\% & $20-100\ M_E$ \\ \hline
191.02 & 0.18 & 2.04 & 6.6\% & $5-18\ M_E$   \\
01     & 1.06 & 11.87& 0.45\%& $0.3-15\ M_J$ \\ \hline
209.02 & 0.68 & 7.62 & 3.4\% & $25-150\ M_E$ \\
01     & 1.05 & 11.76& 0.74\%& $0.3-15\ M_J$ \\ \hline
877.01 & 0.23 & 2.63 & 8.6\% & $6-40\ M_E$   \\
02     & 0.21 & 2.34 & 11\%  & $5-25\ M_E$   \\ \hline
896.02 & 0.28 & 3.14 & 5.9\% & $9-30\ M_E$   \\
01     & 0.38 & 4.26 & 4.9\% & $10-40\ M_E$  \\
\enddata
\tablecomments{The stated fractional uncertainty in the planet radius comes from the radius ratio of the planet and star.  The stars themselves have roughly a 20\% fractional uncertainty in their radii which has not been included here.  The uncertainty in the range of planetary masses is dominated by theoretical uncertainties.
}
\end{deluxetable}

Under these limitations, we estimate a mass range for each object using theoretical models, that are consistent with this level of observational uncertainty \citep{vale2006,vale2007,fort2007,seag2007,gras2009}.  These models of planetary interiors cover a wide range of physical constitutions---from pure hydrogen to pure iron planets.  Obviously, there would be extremes that could not arise in nature.  One can constrain masses based on pure iron and pure water super-Earths, arguing that planet formation scenarios would not allow for such pure constitutions \citep{vale2007,marc2010a,marc2010b}.

In primary planet formation of any flavor, giant impacts and late water delivery are the only plausible way to ``purify'' an initially mixed-materials formation in a protoplanetary disk.  For example, the iron-enhanced bulk composition of Mercury is explained by an early head-on impact with a similar body.  \citet{marc2010a} find that a mass-dependent limit on final mean density (hence, radius) should exist for super-Earth planets more massive than 1 $M_E$, which is significantly less dense than pure iron.  On the low-density bound (high radius), \citet{marc2010b} show that more than about 75\% by mass enrichment in pure water is not possible, but here the upper envelope is not easily constrained due to the possible addition of a H/He envelope and/or extended atmosphere for a hot planet \citep{adam2008,roge2010}.

Starting with the small-size objects, KOI 191.02, with $R_p = 2.0 R_E$, may well be a super-Earth.  It is half the size of the ice giant Uranus ($4 R_E$) and 30\% smaller than GJ1214b ($2.7 R_E$).  However, the mass range of $5-18 M_E$ spans the range between a water-rich world and an iron-rich remnant of a giant impact collision \citep{marc2010a}.  We use a mass of $M_p = 10 M_E$ for KOI 191.02.

Next we have KOI 877.01 and 02 have radii of $2.6 R_E$ and $2.3 R_E$, respectively.  These planets are near the transition to the ice giants Uranus and Neptune, but may be volatile-rich sub-Neptunes or super-Earths like GJ1214b \citep{char2009}.  The estimated mass range is $6-40 M_E$ for KOI 877.01 and $5-25 M_E$ for KOI 877.02.  Since the high-mass, high-density limits are difficult to explain by existing planet formation scenarios we use estimates of $15$ and $10 M_E$, respectively.

Three others, KOI 152.02/03 and KOI 896.02, have sizes similar to Neptune, so we assign them $15 M_E$, noting the large possible range of $9-30 M_E$ and the anticipated transition to planets possessing larger H/He envelopes.  This transition is the reason why KOI 896.01 is estimated at only $M_p = 20 M_E$ despite being much larger than KOI 896.02.  The remaining 4 objects, if confirmed, are likely gas giant planets.  KOI 152.01 ($6.7 R_E$) and 209.02 ($7.6 R_E$) have radii smaller than Saturn ($9 R_E$, $95 M_E$), hence we assign them a smaller mass of $60 M_E$.  They could be more massive than Saturn with large cores (e.g., HD149026b) or lower mass objects with small cores that are dominated by a H/He envelope; the plausible mass range is consequently wide.  The other two, KOI 191.01 and 209.01, have Jupiter sizes and we assign them Jupiter masses noting the cautionary tale of HAT-P-12b and CoRoT-3b above, which also have Jupiter sizes.


\section{Dynamical interactions\label{TTV}}

In these data, we detect no significant TTV signal given the short time baseline.  However, using the measured periods and estimates of the planetary masses from Section~\ref{physparams}, we conduct a Monte Carlo study to determine what TTV signals we expect from these systems with more data.  For this study we assume coplanar orbits; since inclination affects the TTV signal at second order, these results apply to systems with $\phi \lesssim 0.1$ radians, where $\phi$ is the mutual inclination of the planets.

\subsection{Monte Carlo outline}

The masses and periods of the planets used for the Monte Carlo study are fixed to the values shown in Table~\ref{systable}.  Since the short-term TTV signal ($\delta t$) scales in a known manner with planet mass ($\delta t \sim m_{\text{pert}}$ for non-resonant systems and $\delta t \sim m_{\text{trans}} /(m_{\text{trans}} + m_{\text{pert}})$ for resonant systems) it is straightforward to adjust these results for other planetary masses.  Here we define $\delta t$ to be the Root Mean Square (RMS) of the residuals after subtracting a linear ephemeris.

\begin{deluxetable}{cccccccc}
\tablecolumns{8}
\tablewidth{0pc}
\tablecaption{Orbital and physical properties of the systems used in the Monte Carlo study in order of increasing orbital period for each system (not necessarily in order of candidate identification number).\label{systable}}
\tablehead{
\colhead{System} & \colhead{Stellar Mass} & \colhead{Mass 1} & \colhead{Period 1} & \colhead{Mass 2} & \colhead{Period 2} & \colhead{Mass 3} & \colhead{Period 3} \\
\colhead{} & \colhead{($M_\odot$)} & \colhead{} & \colhead{(Days)} & \colhead{} & \colhead{(Days)} & \colhead{} & \colhead{(Days)}
}
\startdata
152 & 1.4 & 15 $M_E$ & 13.5 & 15 $M_E$ & 27.4 & 60 $M_E$ & 51.9$^\star$ \\
191 & 0.9 & 10 $M_E$ & 2.42 & 1  $M_J$ & 15.4 &  & \\
209 & 1.0 & 60 $M_E$ & 18.8 & 1  $M_J$ & 49.3$^\star$ &  &  \\
877 & 1.0 & 15 $M_E$ & 5.96 & 10 $M_E$ & 12.0 & & \\
896 & 0.8 & 15 $M_E$ & 6.31 & 20 $M_E$ & 16.2 & & \\
\enddata
\tablecomments{
An asterix indicates an orbital period estimated from the duration of a single transit event.
}
\end{deluxetable}

Parameters that were adjusted in this Monte Carlo study include the eccentricity of both planets, the longitudes of pericenter, and the mean anomalies at the initial time.  The eccentricities were chosen from a mixture of an exponential and a Rayleigh distribution~\citep{juri2008,zaka2010}:
\begin{equation}
ecc(x) = \alpha \lambda e^{-\lambda x} + (1-\alpha)\frac{x}{\sigma_e^2}e^{-x^2/2\sigma_e^2}
\end{equation}
where $\alpha = 0.38$ gives the relative contributions of the two probability density functions, $\lambda = 15$ is the width parameter of the exponential distribution, and $\sigma_e = 0.17$ is the scale parameter of the Rayleigh distribution.  The value of $\sigma_e$ was found by fitting the distribution of eccentricities in known multiplanet systems measured from radial velocity surveys using only systems with measured eccentricities.  The value of $\lambda$ accounts for the planets found to be in or near circular orbits.  The values of the longitudes of pericenter and mean anomalies were chosen from a uniform distribution rather than using the observed relative longitudes of the various planets in the systems.  Thus, these results are applicable to general systems with the measured period ratios---including the five systems presented here.  For each system, a large sample of initial conditions was generated.  Each realization was integrated for seven years, the time baseline of an extended \textit{Kepler} mission.  The transit times of the planets in the system were tabulated and a linear ephemeris was subtracted.  Finally, the RMS value of the timing residuals was recorded.

A full-scale investigation of the stability of the systems used in the study was not feasible due to the computational cost.  Nevertheless, a few simple criteria were used to eliminate systems that are most likely to be unstable.  First, if any two planets came within two Hill radii of each other (twice the sum of the Hill radii of the two planets) then the system was rejected.  Second, if the semimajor axis of any planet changed by more than 20\% from its initial value then the system was rejected.  Third, if the resulting TTV signal for a given planet was larger than the period ratio of the planets (greater than unity) times the period of the planet then the system was rejected (\citet{agol2005} showed that a TTV signal of order the period of the planet is possible, but that occurs only in the most favorable configuration of a 2:1 MMR and a very massive perturbing planet).  We note that this third criterion eliminated only a small fraction ($\ll 1\%$) of the systems under consideration.  Table~\ref{paramtable} indicates the number of orbits of the innermost planet used in the study, the initial number of systems used in the study, and the final number of systems that satisfied the stability critera above.

\begin{deluxetable}{rccc}
\tablecolumns{4}
\tablewidth{0pc}
\tablecaption{Information about the number of systems used in the Monte Carlo study. \label{paramtable}}
\tablehead{
\colhead{System} & \colhead{Inner Orbits} & \colhead{Initial Systems} & \colhead{Final Systems}
}
\startdata
152 &  185 & 200000 & 3477  \\
191 & 1033 & 15000  & 9069  \\
209 &  133 & 15000  & 12732 \\
877 &  208 & 15000  & 14701 \\
896 &  392 & 15000  & 9566  \\
\enddata
\end{deluxetable}

\subsection{Monte Carlo results}

Here we present some of the results from this study and give an estimate for the expected TTV signal for the five systems considered.  We use as an example the KOI 896 system and then show the essential outcomes for all five systems.  Additional information about the results of the Monte Carlo results is found in the Appendix.

\begin{figure}
\includegraphics[width=0.45\textwidth]{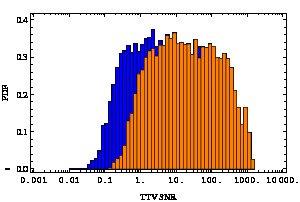}
\caption{Distribution of the TTV signal for the KOI 896 system.  The blue histogram is for the interior planet and the orange is for the exterior planet.  Both histograms overlap for much of the domain.\label{ttv896}}
\end{figure}

The KOI 896 system has two Neptune-size planets just outside the 5:2 mean motion resonance.  Figure~\ref{ttv896} shows the distribution of TTV signals from the simulation in terms of the signal-to-noise ratio (SNR)---that is, the ratio of the size of the TTV signal to the mean timing uncertainty obtained from our pipeline measurements of individual transit times.  The likely minimum size of this TTV signal indicates that additional data should show such deviations from a constant period except in a limited set of configurations.  For example, the fifth percentile of the distribution still gives a TTV signal of a few minutes and has an SNR near unity.  In addition, the fact that it is not precisely situated in an MMR might allow this system to be characterized by the analytic methods of \cite{nesv2008} rather than the full numerical simulations needed for resonant or near-resonant systems.

The width of the resulting TTV distribution can be understood as follows.  
The left-most edge (smallest TTV signal) in Figure~\ref{ttv896} can be estimated from a circular case, employing equation (31) from \citet{agol2005}, extrapolating (and simplifying) from the 2:1 MMR.  Here we would expect TTV signals for the inner planet of order $\delta t \sim m_{\text{pert}}/m_* (P_1/P_2) P_{\text{trans}} \sim 30 \ \text{sec}$.  The TTV signal grows substantially with increased eccentricity.  
The largest expected TTV signal should be of order the period of the planet in question; for KOI 896 this is about 10$^5$ seconds.  These two bounds on the TTV signal are apparent in the results of the Monte Carlo simulation.  Aside from very few realizations that have TTV signals less than 10 seconds, the simulation yields results consistent with the bounds mentioned above.  The median expected signal is roughly 2600 seconds for the inner planet and 6700 seconds for the outer; the ratio of these two values is near their period ratio, also as expected.

\begin{figure}
\includegraphics[width=0.45\textwidth]{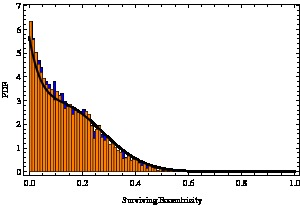}
\caption{Eccentricity distributions for the systems that survived the stability criteria for the KOI 896 system.  The blue histogram is for the interior planet while the orange is for the exterior planet.\label{ecc896}}
\end{figure}

Figure~\ref{ecc896} shows the distribution of eccentricities that survive the stability criteria.  Also shown is the initial distribution for the eccentricities---the solid curve.  While the surviving distribution of eccentricities tracks the initial distribution, one can see that systems with lower eccentricities are somewhat more likely to survive than larger ones.  Note that this figure does not show the final eccentricities, rather it shows the initial eccentricities of the systems that pass all of the stability criteria.

Table~\ref{ttvtable} shows the median, fifth, and 95$^{\text{th}}$ percentiles for the expected TTV signal and SNR for each of the candidate planets.  Histograms similar to figures \ref{ttv896} and \ref{ecc896} for each system are found in the Appendix.  With the possible exception of KOI 191, each of the planets in these five systems will likely have observable transit timing variations by the end of an extended \textit{Kepler} mission.  For KOI 191, even if the TTV signal is small it may yet be detectable simply because there will be a large number of transits (more than 1000) over the duration of an extended mission which may compensate for the low signal-to-noise ratio of the TTV signal to the transit time uncertainties.

\begin{deluxetable}{rrrrr}
\tablecolumns{5}
\tablewidth{0pc}
\tablecaption{Mean timing precision ($\sigma_T$), median and quantile TTV signals expected for each planet in the five systems.  Signal-to-noise ratios are given in parentheses.\label{ttvtable}}
\tablehead{
\colhead{Candidate} & \colhead{$\sigma_T$} & \colhead{Median TTV} & \colhead{5\% TTV} & \colhead{95\% TTV} \\
\colhead{} & \colhead{(sec)} & \colhead{sec/(SNR)} & \colhead{sec/(SNR)} & \colhead{sec/(SNR)}
}
\startdata
152.03 & 790 & 2590  (3.3)& 223  (0.3) & 36100  (45)\\
02     & 700 & 39500 (56) & 4090 (5.8) & 387000 (550)\\
01     & 200 & 18900 (97) & 1830 (9.4) &285000 (1500)\\ \hline
191.02 & 1700& 29 (0.02)  & 6  (0.003) & 5850 (3.4)  \\
01     & 70  & 20 (0.3)   & 9  (0.1)   & 3850 (54)  \\ \hline
209.02 & 330 & 2230 (6.8) & 488 (1.5)  & 79600 (240) \\
01     & 151 & 2300 (15)  & 234 (1.5)  & 73100 (480) \\ \hline
877.01 & 850 & 16500 (19) & 2620 (3.1) & 41900  (49)  \\
02     & 760 & 79100 (100)& 12500 (16) & 200000 (260) \\ \hline
896.02 & 680 & 2630 (3.9) & 103 (0.2)  & 80900 (120)  \\
01     & 370 & 6700 (18)  & 251 (0.7)  & 199000 (540) \\
\enddata
\end{deluxetable}

The primary reason for the small signal in KOI 191 is the large ratio of orbital periods, exceeding 6:1.  Thus, the TTV signal is weakened significantly.  If the outermost planet in KOI 191 were to have an eccentric orbit then 
it would give a periodic TTV signal with a period equal to that of the outer planet as described in Section~4 of \citet{agol2005} \citep[see also][]{bork2003}.

For KOI 209, the expected TTV signal for the inner planet shows an abrupt cutoff and is expected to be larger than a few hundred seconds.  This is because KOI 209.02 has the longest period of all of the inner planets.  Given the time baseline of the extended mission, planets with periods of a few tens of days will likely prove to be among the most interesting for TTV studies as they simultaneously have longer periods (the TTV signal is linear in the period) and will have a sufficient number of transits for a complete analysis.

\begin{figure}
\includegraphics[width=0.45\textwidth]{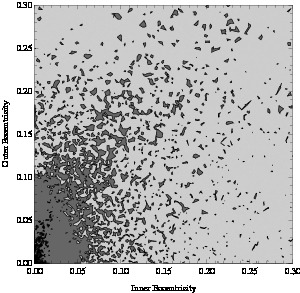}
\caption{Contour plot of the TTV signal for the inner planet in KOI 877 as a function of the inner and outer planet eccentricities.  Note that this is an expanded view of the lower-eccentricity systems, the eccentricity distributions of both planets extend beyond 0.6.  The contours correspond to signal-to-noise ratios of $1\ \text{and}\ 10$ (times of 850 and 8500 seconds).\label{877contours}}
\end{figure}

The proximity of KOI 877 to the 2:1 MMR indicates that this system is likely to have very large variations.  
However, a steep drop in the expected signal occurs when the orbits are nearly circular.  Figure~\ref{877contours} shows an expanded view of the TTV signal for the inner planet in KOI 877 as a function of the inner and outer planet eccentricities.  From this figure one can see that, while the zero eccentricity case exhibits a relatively small TTV signal, eccentricities much larger than 0.01 cause the signal to increase beyond an SNR of unity near $10^3$ seconds ($\sim 15$ minutes).  Should the TTV signal be this size or smaller, it should stringently constrain the eccentricities of both planets in the absence of any other data.  We note that all of these results for the expected TTV signal have significant dependence on the eccentricities of the planets.  One consequence of this fact is that, if a large fraction of these or other multiple systems do not show a TTV signal, then low eccentricity orbits are much more common in multi-planet systems than in single planet systems.

The three-planet system of KOI 152 portends the exciting and challenging studies of systems where there are more than two planets and where multiple planets transit the star.  This system is particularly interesting given the relatively close proximity to the 4:2:1 multibody resonance.  However, it is unlikely that this system occupies this resonance given the estimated orbital periods of the planets---one being estimated from a single transit.  For KOI 152, the middle planet is likely to exhibit the largest TTV signal---being just outside the 2:1 MMR with an interior planet and perhaps just interior to the 2:1 MMR of the exterior planet.

\subsection{Challenges for systems with more than two planets}

One challenge that three-planet systems, such as KOI 152, pose is the confusion that can arise from multiple, competing perturbers in the TTV signal for a particular planet.  We present three broad scenarios for consideration in future studies, although other regimes may exist: 1) nonresonant/nonresonant where there is no mean motion commensurability between any pair of planets, 2) resonant/nonresonant where one pair of planets has a mean motion commensurability while the other does not, and 3) resonant/resonant where any pair of planets lies near a mean motion commensurability.

For the first scenario, the TTV signal due to one perturber should be largely independent of the TTV signal due to the second perturber.  The effect from both perturbers will be of order the perturber to stellar mass ratio, and therefore may be comparable.  But their contributions will contribute linearly to the overall signal and the periodicities in the TTV signal due to one perturber will be independent of the periodicities induced by the other.  In other words, a Fourier transform of the TTV signal would likely show two sets of independent peaks \citep[see][]{stef2006} that can be distinguished provided the data have a sufficient time baseline ($t_{\text{obs}} \gtrsim 1/\Delta f$ where $\Delta f$ is the typical difference in frequency of the most prominent Fourier components between the TTV signals of each planet).

One's ability to identify the orbital elements of the planets in the second scenario, with a resonant pair and a nonresonant companion, will depend upon which planet is transiting.  If a planet in the resonant pair is transiting, then the system is likely to be invertible for two reasons.  First, the resonant signal will be significantly larger (by a factor of the stellar mass to the total planetary mass of the resonating system) and will therefore be more readily identified due to its enhanced signal to noise ratio.  Second, similar to the first scenario, the TTV contribution from the nonresonant perturber will combine linearly with those of the resonant perturber and sufficient data will distinguish their contributions.

Characterizing a system where the nonresonant planet is transiting may be much more challenging.  Here, the transit times will vary on timescales of the libration period of the perturbing, resonating pair of planets.  In general, a signal with a similar period and amplitude may also be generated with a single planet whose orbital period is equal to a multiple of the libration period of the resonating pair.  In addition, resonating systems may exhibit secular evolution on timescales of only a few years, which could also affect the nonresonant planet in a measurable way.

The third scenario, where the system has multiple pairs of planets near MMR, may present serious challenges if only one planet transits.  In favorable configurations, sufficient data will allow one to identify the dynamical interactions among various planets.  This may be more feasible when the system architecture, while resonant, is also hierarchical.  Consider, for example, a 1:2:8 hierarchy where the 4:1 resonance between the outer two planets is likely to be much stronger than the associated 8:1 resonance between the outermost and innermost planets.  On the other hand, compact and strongly interacting systems such as 1:2:4 or 1:2:3 may produce TTV signals that mimic single perturbing systems that are in a different resonance altogether (e.g., a 3:2 or 4:3 MMR).

All of the challenges in characterizing multibody systems from the TTV signal that are listed above are lessened significantly when multiple bodies transit the star.  Most importantly, in a multiply transiting system the periods of more than one planet are known---eliminating confusion regarding, for example, which resonance some planets may occupy.  In addition, if there is a non-transiting perturbing planet in a system with two or more transiting planets then its effects will be present, though perhaps very small, in the TTV signal of each of the transiting planets.  Thus, correctly identifying non-transiting perturbing planets will likely be easier in a multiply transiting system, such as those presented here, than in a singly-transiting system.

\begin{figure}
\includegraphics[width=0.45\textwidth]{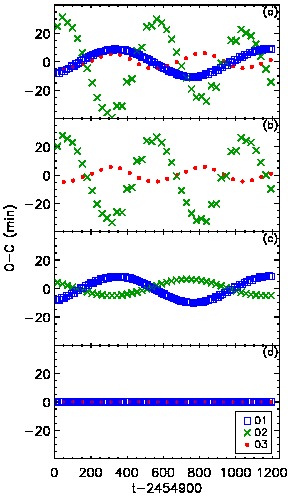}
\caption{Simulated transit times, relative to a linear ephemeris for a system based upon KOI 152.  (a) The full, 3-planet system with parameters given by Table~\ref{152integratorparams}.  (b) The same system, except the inner planet is absent.  The two outer planets interact very similarly to the case with all three planets.  (c) Now the exterior planet is absent.  The inner planet interacts with the middle planet much as in the full system, but the middle planet, now with the dominant interaction of the outer planet absent, oscillates in antiphase with the inner planet.  (d) Now the middle planet is absent.  The inner and outer planets are too widely separated for a significant TTV signal to be present.\label{152omcpanels}}
\end{figure}

We illustrate this point with a realization of the KOI 152 system, given by parameters in Table~\ref{152integratorparams}, starting with circular orbits.  Figure~\ref{152omcpanels} shows the transit timing variations for all three planets and for each pair of two planets separately.  If the inner planet failed to transit, it may be difficult to detect from the transit times of the other planets (compare panels a and b).  However, if the outer planet failed to transit, its presence would be betrayed because the two inner, transiting planets would not oscillate with the same frequency, in antiphase from each other, as they would if they were alone (panel c).  Finally, if the middle planet failed to transit, the large TTVs of both the inner and outer planet would be too large to be due to each other, and the economical hypothesis of an intervening planet could explain both their TTV patterns.

\begin{deluxetable}{ccccc}
\tablecolumns{5}
\tablewidth{0pc}
\tablecaption{Integration parameters for the KOI 152 representative system shown in Figure~\ref{152omcpanels}.\label{152integratorparams}}
\tablehead{
\colhead{Candidate} & \colhead{$P$ (Days)} & \colhead{$\lambda$ (deg)} & \colhead{$e$} & \colhead{$m$ (Earth)}
}
\startdata
03 &13.48  &     -12.551885 &    0.00    &    15 \\
02 &27.40 &       71.211695  &   0.00   &     15 \\
01 &51.94&       -74.197154   &  0.00  &      60 \\
\enddata
\tablecomments{
The initial epoch is t-2454900 = 120 and the assumed stellar mass is 1.4 $M_\odot$
}
\end{deluxetable}

\section{Discussion\label{secDiscuss}}

We presented five \textit{Kepler} targets, each of which has multiple transiting exoplanet candidates.  These candidates have not been fully vetted and therefore none is a confirmed exoplanet.  Yet, each of these systems have passed several important validation steps that are used to eliminate false positive scenarios.  It is difficult to construct viable astrophysical solutions that are consistent with all of the data.  Of particular importance is the fact that many common false positives for singly transiting systems are not viable false positives for multiply transiting systems.  For example, dynamical stability precludes triple star systems with orbital period ratios of order unity.  Another possibility is a foreground star with two background eclipsing binaries, which requires two background eclipsing binaries to be blended within the same point-spread-function (PSF).  Finally, it is unlikely that two systems that have periodic planetary transit features will have period ratios very near two-to-one.

Additional data, such as high precision radial velocity measurements, may provide sufficient evidence to confirm these as planets or refute them as instances of a false positive signal.  At this time these systems are not a sufficiently high priority for the \textit{Kepler} team to conduct such observations.  If these systems are planets and they follow the observed mass-radius relationship of known planets, the \textit{Kepler} mission is likely to find TTV signatures due to their mutual interactions with the possible exception of KOI 191.  Over the course of the mission, a detailed TTV analysis of these systems can constrain their libration amplitudes, masses, and eccentricities---confirming them as planets without the need for many high-precision RV measurements (as in \textit{Kepler}-9 \citep{holm2010}).  Additionally, TTV measurements from the transits of multiple planets can provide better estimates of the stellar properties (e.g., density, limb darkening model) than systems with only a single transiting planet.

The possible observation of Transit Duration Variations (TDVs) in these systems may identify orbital precession, moons, or Trojan companions \citep{ford2007,kipp2009b,kipp2009a}.  For some multiple transiting systems, it should be possible to constrain the relative orbital inclination based on TTVs, TDVs, and the constraint of orbital stability.  Once a sizable and minimally-biased population of multiple transiting systems has been identified (rather than this small and select sample), it will also be possible to characterize the frequency of multiple planet systems, the distribution of mutual orbital inclinations, and identify the orbital architectures of planetary systems.  Collectively, this information will provide considerable insight into the formation, migration, and dynamical evolution of planetary systems.

\appendix

\section{Appendix}

In this appendix we present the balance of the results from the Monte Carlo study of the TTV effect for these systems.  In particular, we start with KOI 877, then 209, 191, and finally the three-candidate system of 152.

\subsection{KOI 877}

\begin{figure}
\includegraphics[width=0.45\textwidth]{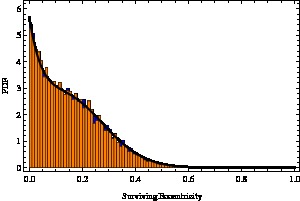}
\qquad
\includegraphics[width=0.45\textwidth]{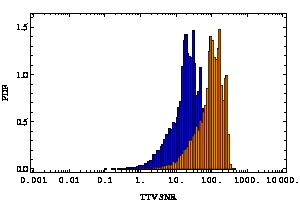}
\caption{Left: Distributions of eccentricities that survive the stability criteria for the KOI 877 system.  Right: Distributions of the TTV signal for the KOI 877 system.  The blue histogram is for the interior planet while the orange is for the exterior planet.\label{results877}}
\end{figure}

The KOI 877 (Figure~\ref{results877}) system lies very near, or possibly within, the 2:1 MMR; and therefore will likely have a large TTV signal.  The Monte Carlo study shows this as the median TTV signal from this system is several hours.  Interestingly, the systems that survive the stability criteria have a virtually identical eccentricity distribution to the initial distribution.  Given the proximity to MMR, it is likely that an independent, in-depth study of the long-term dynamics would reject more of the high eccentricity systems and the expected TTV signal would decline somewhat.  Nevertheless, a system with nearly equal masses near a 2:1 MMR is an ideal scenario to find a TTV signal---even at zero eccentricity.

\subsection{KOI 209}

\begin{figure}
\includegraphics[width=0.45\textwidth]{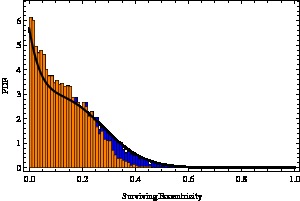}
\qquad
\includegraphics[width=0.45\textwidth]{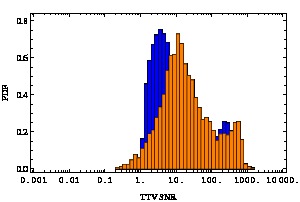}
\caption{Left: Distributions of eccentricities that survive the stability criteria for the KOI 209 system.  Right: Distributions of the TTV signal for the KOI 209 system.  The blue histogram is for the interior planet while the orange is for the exterior planet.\label{results209}}
\end{figure}

For KOI 209, the smaller inner planet is likely to show the largest TTV signal.  Interestingly, the surviving eccentricity distribution of the more massive outer planet is very skewed toward smaller values while the eccentricity of the inner planet more closely matches the initial distribution.  The surviving eccentricity distributions of the KOI 209 system (Figure~\ref{results209}) strongly favor a low-eccentricity outer planet.  Consequently, the expected TTV signal for the outer planet has a tail toward lower values.  This is typical of systems where the planets are not near a mean motion resonance, but interact on secular timescales to exchange significant angular momentum.  The larger planet mass and semimajor axis means that a similar eccentricity results in a much larger angular momentum deficit for the outer planet.  The maximum eccentricity of the inner planet over a secular timescale can be more sensitive to the initial eccentricity of the outer planet than its own initial eccentricity.

\subsection{KOI 191}

\begin{figure}
\includegraphics[width=0.45\textwidth]{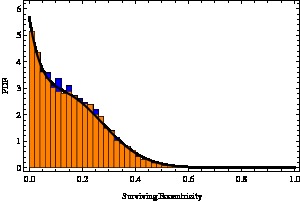}
\qquad
\includegraphics[width=0.45\textwidth]{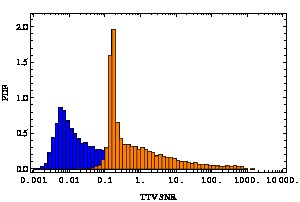}
\caption{Left: Distributions of eccentricities that survive the stability criteria for the KOI 191 system.  Right: Distributions of the TTV signal for the KOI 191 system.  The blue histogram is for the interior planet while the orange is for the exterior planet.\label{results191}}
\end{figure}

The hierarchical structure of the KOI 191 system (Figure~\ref{results191}) means that a TTV signal is likely to be significant only if there is significant eccentricity in the orbit of either planet.  The sharp peak in the TTV distribution of the outer planet near 10 seconds agrees with the expectation for an outer transiting planet in a system where the planet interactions are negligible (section~3 of \citet{agol2005}).  In this scenario the outer planet probes the astrometric deviations of the star due to the inner planet.  A rough estimate of the size of this effect can be found by taking the characteristic displacement of the star from the barycenter and divide by the characteristic velocity of the outer planet $(a_1 m_1/m_*)/(a_2/P_2)\simeq 10$sec.  It may be possible to observe variations in the duration of the transit of the outer planet due to the movement of the star within the inner binary while the outer planet is transiting in other hierarchical systems, but perhaps not with the KOI 191 system.

\subsection{KOI 152}

For this system, all three planets should have a sizeable TTV signal due to the proximity of the 2:1 MMR between the inner and outer pair.  As mentioned above, the middle planet will likely show the largest TTV signal.  One would expect the TTV signal for the outer planet to be roughly half that of the middle planet---it has a factor of two larger period, but the perturbing, middle planet is only one quarter the mass.  The eccentricity distribution of the outer two planets favor smaller eccentricities.

\begin{figure}
\includegraphics[width=0.45\textwidth]{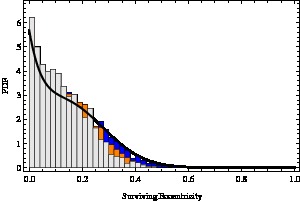}
\qquad
\includegraphics[width=0.45\textwidth]{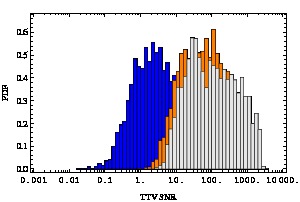}
\caption{Left: Distributions of eccentricities that survive the stability criteria for the KOI 152 system.  Right: Distributions of the TTV signal for the KOI 152 system.  The blue histogram is for the interior planet, orange is for the middle planet, and light gray is for the exterior planet.\label{results152}}
\end{figure}

\bibliographystyle{plainnat}

\end{document}